# Improving Editorial Workflow and Metadata Quality at Springer Nature

Angelo A. Salatino[1], Francesco Osborne[1], Aliaksandr Birukou[2], Enrico Motta[1]

[1] Knowledge Media Institute, The Open University, MK7 6AA, Milton Keynes, UK
{angelo.salatino,francesco.osborne,enrico.motta}@open.ac.uk
[2] Springer-Verlag GmbH, Tiergartenstrasse 17, 69121 Heidelberg, Germany
aliaksandr.birukou@springer.com

**Abstract.** Identifying the research topics that best describe the scope of a scientific publication is a crucial task for editors, in particular because the quality of these annotations determine how effectively users are able to discover the right content in online libraries. For this reason, Springer Nature, the world's largest academic book publisher, has traditionally entrusted this task to their most expert editors. These editors manually analyse all new books, possibly including hundreds of chapters, and produce a list of the most relevant topics. Hence, this process has traditionally been very expensive, time-consuming, and confined to a few senior editors. For these reasons, back in 2016 we developed Smart Topic Miner (STM), an ontology-driven application that assists the Springer Nature editorial team in annotating the volumes of all books covering conference proceedings in Computer Science. Since then STM has been regularly used by editors in Germany, China, Brazil, India, and Japan, for a total of about 800 volumes per year. Over the past three years the initial prototype has iteratively evolved in response to feedback from the users and evolving requirements. In this paper we present the most recent version of the tool and describe the evolution of the system over the years, the key lessons learnt, and the impact on the Springer Nature workflow. In particular, our solution has drastically reduced the time needed to annotate proceedings and significantly improved their discoverability, resulting in 9.3 million additional downloads. We also present a user study involving 9 editors, which yielded excellent results in term of usability, and report an evaluation of the new topic classifier used by STM, which outperforms previous versions in recall and F-measure.

## 1 Introduction

Identifying the research topics that best describe the scope of a scientific publication is a crucial task for editors, in particular because the quality of these annotations determines how effectively users are able to discover the right content in online libraries. A high-quality representation of research publications has also an effect on the performance of approaches to discovering and querying scientific articles [1], producing smart analytics [2], detecting research communities [3], extracting research entities [4], recommending publications [5], forecasting research topics [6], and so on.

Springer Nature (SN), the world's largest academic book publisher, produces for each new book a high quality list of relevant topics, which are used for describing the book content in the metadata. This task is particularly complex in the case of books covering conference proceedings, as these may easily contain over 100 different contributions, each of which may be relevant to several areas of research. As a result, the set of topics covered by the proceedings of a conference can be very large and it is not trivial to manually select a small number of topics that best describe the entire set of contributions. In particular, it is easy to miss the emergence of a new topic or to assume that some topics are still popular when this is no longer the case. For these reasons, this task has been entrusted to senior editors, which typically select a list of topics on the basis of their own expertise in the field, a visual exploration of titles and abstracts, and, optionally, a list of keywords derived from the call for papers of the conference in question. Hence, this process has traditionally been very expensive, time-consuming, and confined to a few expert editors. Moreover, the resulting topics may vary according to the background of the editor and the same topic could be referred to by means of different labels (e.g., LOD, Linked Data) or at a different abstraction level (e.g., Deep Learning, Machine Learning).

For these reasons, in 2016 we developed Smart Topic Miner [7], an application supporting Springer Nature editors in annotating publications in terms of a set of topics drawn from a large ontology of research areas in Computer Science [8]. Since then STM has been adopted by editors in Germany, China, Brazil, India, and Japan to annotate all book series covering conference proceedings in Computer Science, for a total of about 800 volumes per year. Over the past three years, STM has iteratively evolved in response to feedback from the users and has been extremely successful in both reducing costs and improving the quality of the metadata. It has drastically reduced the time used to annotate proceedings, from about 30 to 10-15 minutes for each volume, and allowed the task to be performed by junior editors or editorial assistants, ultimately achieving an overall 75% cost reduction. The resulting metadata have improved significantly the discoverability of the relevant books, resulting in about 9 million additional downloads. We believe that the evolution of this small prototype to a high impact system adopted by one of the main academic publishers constitutes an exemplary success story regarding the adoption of semantic technologies in large companies.

In this paper we introduce Smart Topic Miner 2 (STM 2), the most recent version of the tool, and describe the evolution of the system over the years, the key lessons learnt, and the impact on the Springer Nature workflow and on the discoverability of the relevant publications. We also present a user study on STM 2 involving 9 editors, which yielded excellent results in term of usability, and report an evaluation of the new research topic classifier, which outperforms the previous versions in terms of both recall and F-measure. The main novelties with respect to the earlier report on this work [7] include: 1) a new approach to identifying research topics and producing an explanation for each suggested topic, 2) a new interactive interface, 3) a new capability for the system to take into account the annotation of previous editions of the conference in question, 4) the integration of STM 2 with the CSO Portal [8] and the SN editorial systems.

The paper is structured as follows. In Section 2, we discuss the evolution of STM and describe the latest version in detail. Section 3 presents the user study and the evaluation of the classifier. Section 4 describes the uptake and impact of STM, and Section 5 discusses the relevant work. Finally, in Section 6 we summarise the main results from this work and discuss our plans for further developing STM in Springer Nature.

## 2  Smart Topic Miner 2

Smart Topic Miner 2 is a web application that assists editors in classifying books and more in general any collection of research papers. Specifically, it takes as input XML files describing the metadata of one or more books and returns:

- A taxonomy of the relevant topics drawn from the Computer Science Ontology (CSO) [8], which is the largest taxonomy of research topics in the field;
- A set of relevant Product Marked Codes (PMCs), Springer Nature internal classification;
- An explanation for each topic, in terms of the text excerpts that triggered the topic identification;
- A list of chapters from the book annotated with topics from CSO.

The editors use an interactive interface to explore this output, check why specific topics were inferred by the system, compare them with the annotations produced in previous editions, and include or exclude specific topics or PMCs according to their expertise. The resulting sets of topics and PMCs are eventually included in the metadata of the publications. These are then used for classifying proceedings in digital and physical libraries and consequently improving the discoverability of the publications in SpringerLink and several other digital libraries and third-party sites. They are also used to power Smart Book Recommender [9], an ontology-based recommender system, which supports the editorial team in selecting the products to market at specific venues.

Figure 1 shows the STM 2 architecture, which consists of four main components: i) the user interface, ii) the parser, which elaborates the input files, iii) the back-end which consists of five sub-components, and iv) the knowledge bases.

A demo version of STM 2 is available at http://stm-demo.kmi.open.ac.uk.

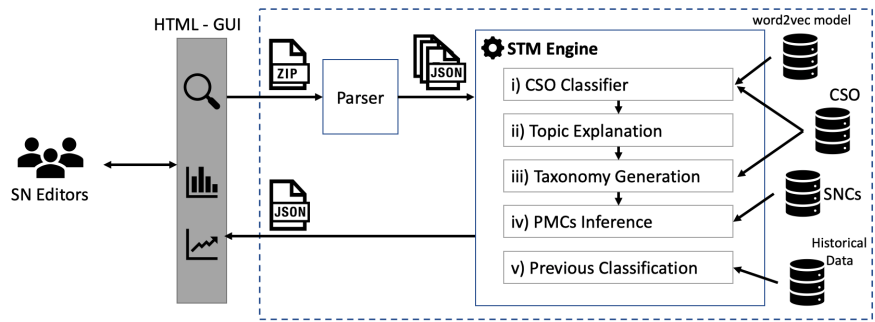

Figure 1. The STM architecture

In the next subsection we summarize the evolution of STM over the last three years, and describe the current version, STM 2, in detail. In Section 2.2 we discuss the knowledge bases used by the system, in Section 2.3 we present the approach to infer research topics and PMCs, and in Section 2.4 we illustrate the user interface.

### 2.1  STM Evolution

Most of the features implemented in STM 2 were designed to address the feedback of the editors using this application since 2016. In this section we summarize the evolution of the system over the last three years by discussing the main changes and their rationale.

*2.1.1 Back-end.*

A fundamental component of the STM back-end is the classifier used for detecting a set of topics for each chapter. In the original implementation, the STM classifier identified the label of the topics from the ontology (e.g., Linked Data) and then inferred all their super topics (e.g., Semantic Web, AI, Computer Science). However, an analysis of the output with SN editors revealed that our method was missing some variations of the labels that were not covered in the ontology. For this reason, we designed a new approach (the CSO Classifier 1.0 [10]) which selects all the ontology topics which have a Levenshtein similarity higher than a threshold with n-grams extracted from the text. A new meeting with the editors confirmed that this problem was solved, but revealed a more subtle issue: some topics were never explicitly mentioned and could be inferred just by considering the text as a whole. For instance, the abstract of a chapter about "online communities" never mentioned this label, a similar variation, or any of its subtopics in the ontology. However, it mentioned several related terms, such as the name of popular social networks and words related to network analysis. Hence, we recently created a new version of the CSO Classifier (2.0 [11], described in Section 2.3.2) that uses NLP and word embeddings to detect also implicit topics. This solution outperforms previous versions in terms of both recall and F-measure (see Section 3.2). Editors have also confirmed that the current output is now much more comprehensive, both at book and at chapter level.

*2.1.2 User interface.*

The user interface of STM 2 (Figure 2) is visibly different from the 2016 version (Figure 3). The first important change regards the way topics are presented. In the original version, the taxonomy was given as text in order to allow the editors to easily export the output. However, editors wanted to be able to interact with the topics for the purpose of requesting an explanation, renaming them, or inserting a missing sub-topic. Hence, we implemented a new interactive interface that displays the taxonomy as a tree and allows editors to right click on the topics to access a number of functionalities.

The second change is the introduction of a new input menu that allows the editors to load several books at once. Indeed, the proceedings of a conference can consist of 2-10 volumes, which need to be analysed together and will share the same topics and PMCs.

The third important change regards the way the editors control the granularity of the representation. Originally, STM used a set covering algorithm that, given a "granularity value" from 1 to 5, would return a more or less comprehensive version of the taxonomy. This solution had two main issues. First, the "granularity value" was arbitrary and there was no straightforward explanation on why a certain topic was included or not. Secondly, the editor needed to submit a new request to the back-end every time they had to change the granularity, which was quite time consuming. In STM 2 we addressed these issues by moving the process to the front-end. Now the back-end produces a taxonomy including all relevant topics, and the interactive interface shows only the ones associated with a minimum number of chapters. The user can change this value with a sliding bar, making the displayed taxonomy more or less inclusive. This solution was greatly appreciated by the editors, since it makes them feel in full control of the filtering mechanism. Interestingly, while the previous version produced arguably better summarizations of the taxonomies, editors prefer this simple solution since it produces a more predictable outcome. This suggests that transparency and understandability are of the utmost importance when supporting human experts in the exploration of automatically generated knowledge bases.

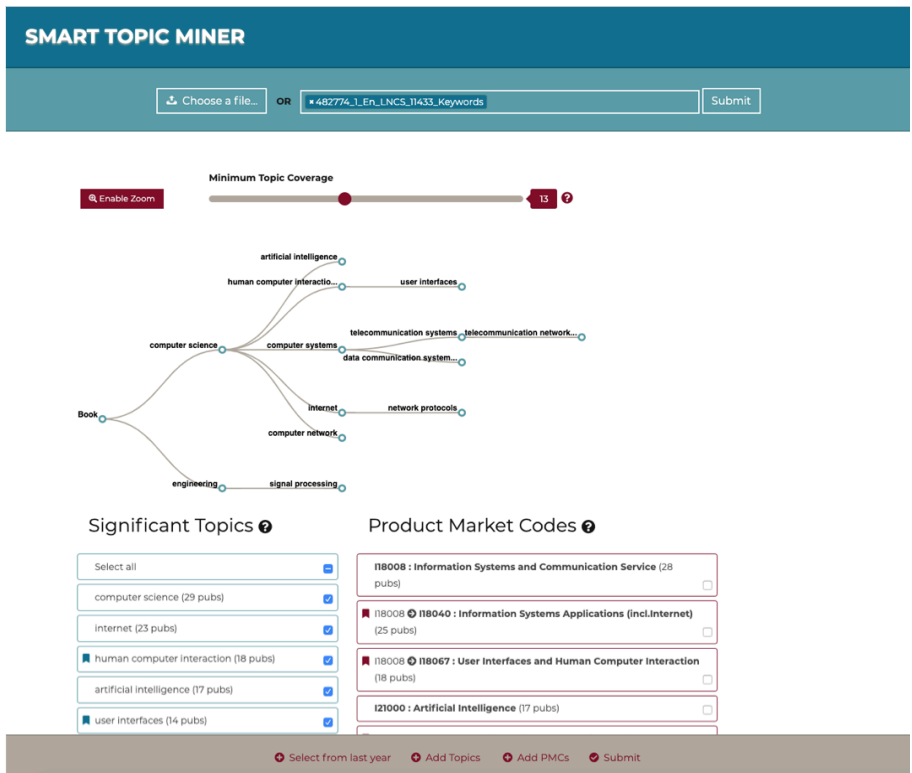

Figure 2. Smart Topic Miner 2.0 interface.

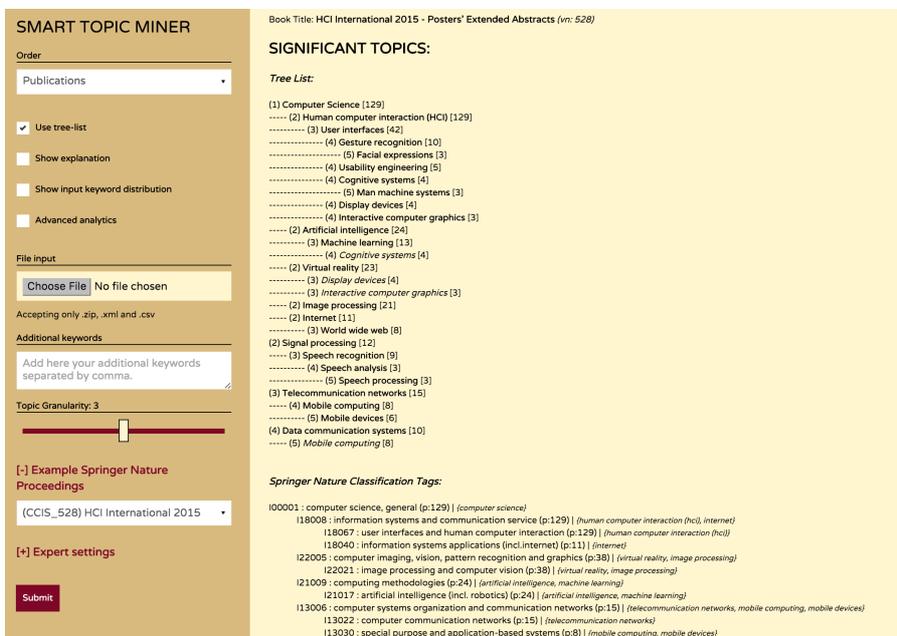

Figure 3. Smart Topic Miner 1.0 interface.

The final change regards the left bar that used to include the main menu and all the options. The editors noticed that even if they did not typically use it after loading the proceedings, it was still taking a lot of screen space, reducing the area available for navigating the taxonomy of topics. More generally, we learnt that our users did not consider several settings that we included in the first version. For instance, many options for changing the parameters of the classifiers were considered too technical and opaque by the users. They preferred a simple and clean interface allowing them to focus on the topic taxonomy. For this reason, we removed the bar and all the superfluous settings and simplified the front-end as much as possible. The resulting interface is described in detail in Section 2.4.

*2.1.3 Comparison with previous year annotations.*

The previous version of STM only took in consideration the specific proceedings book loaded by the editor when producing the set of topics and PMCs. However, the proceeding books of previous editions may have been already annotated. The editors suggested that in some cases it would be easier for them if they could start from the annotations of previous editions, which had already been inspected and verified by an editor, and update them by taking in consideration new research trends for the conference. For this reason, STM now identifies the conference series, retrieves the annotations used for the previous edition, highlights these topics, and allows the editors to select them as default starting points. An additional advantage of this solution is that it clearly shows to the editors the fading or emerging topics in a certain venue. This information supports relevant editorial and marketing decisions, such as the publication of a book on the emerging topics.

*2.1.4 Integration with the CSO Portal.*

Representing relevant topics as a taxonomy has always been one of the most appreciated functionalities of STM, since it provides an intuitive way to explore topics and their relationships. However, this taxonomy offers only a partial view of the original ontology, since it includes exclusively the most significant topics from the input books. Hence, editors had no way to examine the original ontology - e.g., for better understanding the nature of a topic or for assessing how good was the coverage of a certain venue with respect to a research field. A related problem was the lack of an automatic and robust mechanism for reporting potential mistakes in the knowledge base, such as an erroneous *subTopicOf* relationship that was causing incorrect inferences. We addressed these issues by integrating STM 2 with the CSO Portal, a web application that enables users to explore and provide granular feedback on CSO. This solution allows the editors to inspect each topic, read a brief summary extracted from DBpedia, check all related topics, and flag incorrect or dubious relationships.

*2.1.5 Integration with Springer Nature systems.*

The first version of STM was not integrated with the SN editorial system. Therefore, the editors had first to obtain a file with the publication metadata and import it in STM, then copy-paste the resulting set of topics and PCMs in an online form in order to submit them to the editorial systems. This process was time-consuming and not very robust. Hence, we integrated STM 2 with the SN metadata systems and production workflows, so that the editors can now load the metadata and submit the final annotations to the editorial system within the application interface.

## 2.2 Background data

Smart Topic Miner relies on three main knowledge sources: i) the Computer Science Ontology (CSO), ii) the Product Market Codes (PMCs), and iii) the metadata of Springer Nature publications. It also exploits a Word2Vec model trained on about 4.5M publications in Computer Science. We describe all of them in the following subsections.

### 2.2.1 The Computer Science Ontology

The Computer Science Ontology (CSO) [8] is a large-scale ontology of research areas that currently includes 14K topics linked by 162K semantic relationships. CSO is available through the CSO Portal[1], a web application that enables users to download, explore, and provide granular feedback on CSO at different levels. It was generated by running the Klink-2 algorithm [12] on a large dataset of research publications in the field of Computer Science [2]. Since it is regularly updated by re-running Klink-2 on recent corpora and integrating user feedback from the CSO Portal, it includes also emerging topics, which are not typically covered by human crafted taxonomies.

The CSO data model[2] is an extension of SKOS[3]. It includes three main semantic relations: *superTopicOf*, which indicates that a topic is a super-area of another one (e.g., Semantic Web is a super-area of Linked Data), *relatedEquivalent*, which indicates that two topics can be treated as equivalent for the purpose of exploring research data (e.g., Ontology Matching and Ontology Mapping, and *contributesTo*, which indicates that the research output of a topic contributes to another. CSO is licensed under a CC BY 4.0 and is available for download at https://cso.kmi.open.ac.uk/downloads.

### 2.2.2 Classification system at Springer Nature

The Product Market Codes is a three-level mono-hierarchical classification system used by Springer Nature to categorize proceedings, books, and journals. The Computer Science section includes 103 categories characterizing both research fields (e.g., *I23001 – Computer Applications*) and domains (e.g., *I23028 - Computer App. In Social and Behavioral Sciences*). This classification was recently updated by using an innovative ontology evolution approach [13] for selecting a new set of topics that best fit SN catalogue. The resulting identifiers are used in the metadata describing the contents for the Springer Nature website[4] as well as third-party libraries and bookshops.

We integrated CSO and PMCs by means of 332 relationships, so that every PMC is now associated to a set of related CSO concepts, as described in [7]. For instance, we mapped the *computer communication networks* category to CSO topics such as Network Security, Telecommunication Networks, Wireless Telecommunication Systems, Wireless Sensor Networks, and so on.

### 2.2.3 Metadata of Springer Nature publications

STM 2 has access to a database of metadata contains titles, abstracts, keywords and other information describing about 50K books in the field of Computer Science, including 10K proceedings [14]. In this dataset each proceedings book is associated with an ID identifying its conference series, as well as with the topics and PMCs chosen by editors. This information is used by STM 2 to identify the previous edition of a conference and retrieve all relevant data.

---

[1] Computer Science Ontology Portal - https://cso.kmi.open.ac.uk
[2] CSO Data Model - https://cso.kmi.open.ac.uk/schema/cso
[3] SKOS Simple Knowledge Organization System - http://www.w3.org/2004/02/skos
[4] Springer Link - https://link.springer.com

### 2.2.4 Word2Vec model

In order to support the classification of text, we trained a Word2Vec [15] word embedding model on a corpus extracted from the Microsoft Academic Graph (MAG), which is an heterogeneous graph containing scientific publication records, citation relationships, authors, institutions, journals, conferences, and fields of study. Specifically, we considered titles and abstracts of 4,654,062 English publications in the field of Computer Science. We pre-processed these data by replacing spaces with underscores in all n-grams matching the CSO topic labels (e.g., "semantic web" became "semantic_web") and for frequent bigrams and trigrams (e.g., "highest_accuracies", "highly_cited_journals"). These frequent n-grams were identified by analysing combinations of words that co-occur together (collocations), as suggested in Mikolov et al. [15]. This solution allows STM to better disambiguate concepts and treat terms such as "deep_learning" and "e-learning" as completely different words.

More details on how we trained the model and its parameters are available in Salatino et al. [11].

## 2.3 Back-end

The back-end processes one or more proceedings books according to five steps:
1. *Parsing of the metadata*, in which STM 2 extracts the metadata associated to each chapter;
2. *Topic Extraction*, in which STM 2 uses the CSO Classifier to map each publication to a selection of research concepts drawn from CSO;
3. *Generation of Explanations*, in which STM 2 generates for each topic the list of text excerpts that lead to its identification;
4. *Inference of PMCs*, in which the selected topics are used to infer a number of PMCs, using the mapping between CSO concepts and SN codes;
5. *Taxonomy generation and retrieval of previous annotations*, in which STM 2 produces a taxonomy of topics and retrieves the classification of the previous editions.

### 2.3.1 Parsing of the metadata

In the first step, STM 2 extracts all XML files available within the input ZIP files. Then, for each XML file, it retrieves the metadata associated to each chapter: title, abstract, list of keywords, chapter id, volume number, and optionally the conference series identifier. In proceedings books, each chapter is typically a research paper accepted by a conference or a workshop.

### 2.3.2 Topic extraction

At this stage, STM 2 extracts topics from the metadata of a book by running the CSO Classifier [11] on the title, abstract and keywords of each chapter. The CSO Classifier is a tool that we developed for automatically classifying research papers in terms of relevant concepts drawn from CSO. Figure 4 reports its workflow. It identifies topics by means of two different components, the syntactic module and the semantic module, then it combines their outputs and enhances the resulting set by including all relevant super-topics. Its pseudocode is available at https://cso.kmi.open.ac.uk/cso-classifier. In the following we briefly describe it; we refer the interested reader to [11] for additional details.

The *syntactic module* removes English stop words and collects unigrams, bigrams, and trigrams. Then, for each n-gram, it computes the Levenshtein similarity with the labels of the topics in CSO. Finally, it returns all research topics whose labels have similarity to one of the n-grams, which is equal to or higher than a threshold.

The *semantic module* uses part-of-speech tagging to identify candidate terms composed of a proper combination of nouns and adjectives and decomposes them in unigrams, bigrams, and trigrams. For each n-gram, it retrieves its most similar words from the Word2Vec model described in Section 2.2.4. For this task, the n-gram tokens are initially glued with an underscore, creating one single word, e.g., "semantic_web". If this word is not available within the model vocabulary, the classifier uses the average of the embedding vectors of all its tokens. Then, it computes the relevance score for each topic in the ontology as the product between the number of times it was identified in those n-grams (frequency) and the number of unique n-grams that led to it (diversity). Finally, it uses the elbow method [16] for selecting the set of most relevant topics.

The CSO Classifier aggregates the topics returned by the two modules and enriches them by inferring the list of all their super topics, exploiting the *superTopicOf* relationship within CSO [8]. For instance, given the topic "Neural Networks", it will infer "Machine Learning", "Artificial Intelligence", and "Computer Science". This feature allows us to capture both high-level fields and very granular research areas, in order to generate a comprehensive representation of the proceedings books. In order to exclude generic and ambiguous terms (e.g., "language", "learning", "component", etc.), the classifier does not consider the 3,000 most frequent words in the Word2Vec model.

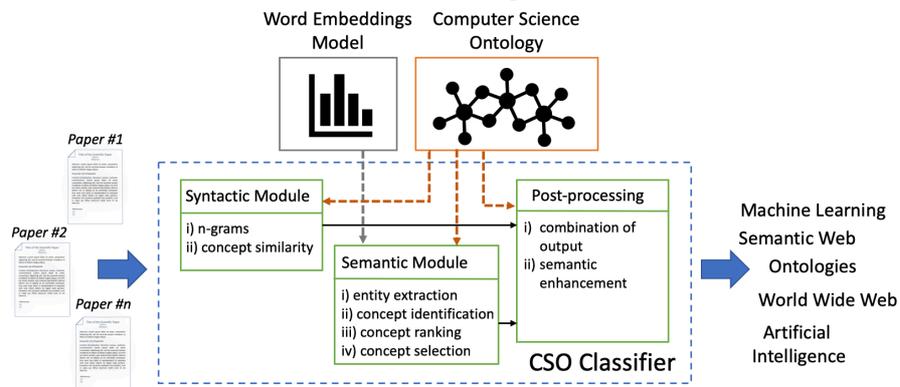

Figure 4. Workflow of the CSO Classifier.

### 2.3.3 Generation of Explanations

Editors do not like to work with a black box. It is critical for them to be able to understand and verify why STM 2 identifies a certain topic. Therefore, STM 2 generates an explanation for each topic in terms of a distribution of text excerpts from which the topics was inferred. This process is important both for building trust in the system and for detecting possible mistakes. In order to produce these explanations, STM 2 first maps the topics with the portions of text that triggered them during the identification. It then associates the text excerpts of a topic to all its super-topics. Finally, it orders them by the number of chapters in which they appear. For instance, the explanation for the topic "Natural Language Processing" will be composed of fragments of texts and their frequencies such as *language processing (6), text mining (6), information extraction (4), keyphrase extraction (4), textual data (3), syntactic analysis (2)*, and so on. These explanations are one of the most appreciated features of STM 2. Indeed, editors reported that in several occasions they assumed that a suggested topic was wrong, but, after checking the explanation, they realized it was indeed addressed in several chapters.

*2.3.4 Inference of PMCs*

In this step, STM 2 uses the mapping between the PMCs and CSO to infer all relevant PMC identifiers. It does so by inferring each PMC that subsumes one of the topics in CSO according to the mapping described in Section 2.2.2. For example, if the Cryptography topic was yielded by the previous step, STM will infer the identifier 'I15033 - Data Encryption' (at the third level), 'I15009 - Data Structures, Cryptology and Information Theory' (second level) and 'I00001 - Computer Science, general' (root). It then associates to each identifier the total number of chapters covered by the associated topics to assist the editors in assessing its significance for the book under analysis.

*2.3.5 Taxonomy generation and retrieval of previous annotations*

Finally, STM 2 builds a taxonomy of topics, using the *subTopicOf* relationships in CSO. It then uses the conference series ID from the metadata to retrieve the topics and PMCs associated to the same conference in the previous year. This information will be displayed alongside the current list of topics and PMCs.

## 2.4 User Interface

The STM 2 interface (Figure 2) is composed by three main components: 1) a top menu for loading the metadata of one or more books, 2) a main panel for inspecting and selecting topics and PMCs, and 3) a bottom menu for submitting the classification and accessing further options.

The main panel also consists of three main parts: 1) a taxonomy of CSO topics, 2) the main menu for selecting topics and PMCs, and 3) a list of chapters from the book. The taxonomy represents topics as nodes linked by their *superTopicOf* relationships from CSO. It includes by default all topics associated with a minimum number of chapters. The editors can control this value by mean of a sliding bar. For instance, in Figure 2 the taxonomy includes all the topics which appear in at least 13 (out of 29) chapters. Topics can be collapsed or expanded and right clicking on one of them opens a contextual menu with several options:

- *Show Explanation*, which displays all relevant text excerpts;
- *Explore in CSO*, which allows editors to examine the topic in the CSO Portal, navigate CSO, and leave a feedback;
- *Rename*, which allows editors to change the label of the topic. For instance, "internet of things" has been renamed as "Internet of Things (IoT)";
- *Remove*, which allows editors to remove the topic from the output;
- *Add subarea*, which allows editors to add a new subarea to the selected topic.

The main menu displays topics and PMCs ordered by frequency and allows editors to select or re-rank them. The topics and PMCs used by the previous edition of the conference are marked with an icon.

The lower part of the main panel displays a summary of all the chapter/articles in the input books. STM 2 shows for each chapter its title, abstract, keywords, and the topics from CSO. It also highlights all text excerpts from the abstracts that triggered the identification of a topic.

The last component of the user interface is the bottom menu, which provides several functionalities that allow editors to interact with the classification outcome and export the final result. The button *Select from last year* allows editors to select all topics and PMCs that were used in the previous classification. The buttons *Add Topics* and *Add PMCs* allow the manual insertion of a topic or a PMC. All the edits are recorded by STM 2 and will be considered when generating a new version of CSO. Finally, the button *Submit* sends all the selected topics and PMCs to the editorial system.

## 3 Evaluation

In this section we discuss the results of a user study on STM 2 and report an evaluation of the classifiers used by STM over the years.

### 3.1 User Study

We performed a qualitative study on STM 2 to assess the quality of its output, the clarity of the explanations, the impact on the editor workflows, and the usability of the user interface[5]. To this end, we organized individual sessions with nine SN editors from Heidelberg, São Paulo, and Beijing. They had on average 4.8 years of experience as editors, and three out of nine had at least 7 years of experience. All of them claimed to have wide knowledge of the research topics in their fields, and seven have a significant knowledge of Springer Nature Classification. Three of them considered themselves also experts at working with digital proceedings.

We demoed STM 2 showing the new functionalities for about 30 minutes and then asked them to use the application for classifying several proceedings in their fields of expertise for about 30 minutes. We took advantage of this session to gather further feedback about new potential use cases. After the hands-on session the editors filled a three-parts survey about their experience. The first part assessed the editor background and expertise, the second part included seven open questions, and the third part was a standard System Usability Scale (SUS)[6] questionnaire to assess the usability of the application. Here we summarize their answers to the open questions.

**Q1. How do you find the interaction with the STM interface?** Four editors considered it "easy" to use, one of them found it "user friendly", two other editors were positive about it. Two editors found some minor issues: the first suggested that the explanatory tooltips could be made more readable, whereas the second argued that he would prefer to load the proceedings by filling in the acronym and the year of a conference (e.g., ISWC 2018) rather than the volume number (e.g., LNCS 11136 and 11137).

---

[5] The data collected during this evaluation are available for download at http://doi.org/10.21954/ou.rd.7951496.

[6] System Usability Scale (SUS) - https://www.usability.gov/how-to-and-tools/methods/system-usability-scale.html

**Q2. How effectively did STM support you in classifying books/publications?** All the editors stated that the application had an extremely positive effect on their work, commenting that it was "quite effective", "very good tool", "extremely helpful", "useful", and so on.

**Q3. What were the most useful features of STM?** The most useful features included: the ability to explore topics at different granularities (five editors), the ability to automatically extract the list of topics and the list of PMCs (four editors), and the possibility to see previous conference classifications (one editor).

**Q4. What are the main weaknesses of STM?** Editors did not flag any particular weakness. Two editors pointed out that the main weaknesses of the previous version had been fixed. One of them pointed out that weaknesses might appear when used extensively.

**Q5. Can you think of any additional features to be included in STM?** The suggested features were: 1) the ability to suggest the primary PMC in the proposal phase, when the title and abstract of the chapters are not yet available (three editors), 2) the ability of saving a classification as work-in-progress and resume it later (two editors), and 3) the ability to see the rank of PMCs of the previous year classification (two editors).

**Q6. How comprehensive/accurate do you consider the list of topics returned by STM?** Six editors found the list of topics very accurate and comprehensive. Three junior editors claimed that they were not yet confident in assessing the quality of the outcome by themselves and usually relied on senior editors for a final confirmation before submitting.

**Q7. How comprehensive/accurate do you consider the list of PMCs returned by STM?** The answers were very similar to those for Q6. As before, six editors found the list very accurate, while three junior editors admitted they preferred consulting with senior editors to verify the quality of the outcome.

The SUS questionnaire confirmed the good opinion of the editors, scoring 82.8/100, which is equivalent to an A grade and places STM 2 in the 93% percentile rank[7]. Considering that the 2016 version obtained 76.6/100, equivalent to a B grade, this result confirms that the changes applied over the years, responding to the editors' feedback, successfully increased the usability of STM. All editors felt very confident in using STM (with an average score of 4.3±0.5) and thought that STM was easy to use (4.4±0.72). In addition, they were happy to use STM frequently (4.6±0.5) and did not think that it was complex to use (1.6±0.7) or that they would need the help of a technical person to use it in the future (1.8±0.6).

### 3.2 Classifier evaluation

As discussed in Section 2.1.1, STM has used three classifiers in its lifecycle: 1) the STM Classifier in 2016, 2) the CSO Classifier 1.0 in 2018, and 3) the CSO Classifier 2.0 since 2019. The STM classifier, described in [7], identifies the label of the topics in the text. The CSO Classifier 1.0, described in [10], selects topics having a Levenshtein similarity higher than a threshold with a set of n-grams extracted from the text. The CSO Classifier 2.0, described briefly in Section 2.3.2 and more extensively in [11], uses

---

[7] Percentiles of SUS - https://measuringu.com/interpret-sus-score/

NLP and word embeddings to identify also topics that are not explicitly mentioned in the text.

We evaluated these three methods on a gold standard of 70 research papers[8], each of them annotated by three domain experts. Table 1 shows the results of the evaluation. The STM classifier yielded the best performance in term of precision (80.8%), but the worst in term of recall. The CSO Classifier 1.0 obtained a marginally lower precision, but a higher recall. Naturally, the good precision of these methods derives from the fact that they focus on topics that are explicitly mentioned in the text. The CSO Classifier 2.0 outperformed the other solutions in both recall (75.3%) and F-measure (74.1%). The loss in precision is not so important in the context of STM for two main reasons. First, when aggregating the results from multiple publications (in several cases more than 100 papers) errors tend to cancel each other. Secondly, the resulting topics are manually checked by the editors, who prefer receiving a more comprehensive set and filtering out some mistakes rather than missing some interesting (and possibly emerging) topics.

Table 1. Precision, recall, and F-measure of the classifiers. In bold the best results.

| Classifier | Precision | Recall | F-Measure |
|---|---|---|---|
| C1: STM 2016 (STM Classifier) [7] | **80.8%** | 58.2% | 67.6% |
| C2: STM 2018 (CSO Classifier 1.0) [10] | 78.3% | 63.8% | 70.3% |
| C3: STM 2019 (CSO Classifier 2.0) [11] | 73.0% | **75.3%** | **74.1%** |

## 4  Uptake and Impact

STM was introduced in Springer Nature in 2016 and has since been used routinely by their Computer Science editorial team to annotate all book series covering conference proceedings in Computer Science, including LNCS, LNBIP, CCIS, IFIP-AICT and LNICST, for a total of about 2,400 volumes over the last three years.

The adoption of STM has brought three main benefits. First, it halved the time needed for classifying a proceedings book from 30 to 10-15 minutes, saving more than 600 working hours since its introduction. Second, it reduced the complexity of this task, which traditionally was performed only by Senior Editors with a vast experience in the relevant research fields. This in turn allowed to entrust the annotation of new volumes to junior editors and editorial assistants, distributing the workload in the editorial team, freeing up the time of the senior editors. As a result, the overall cost of the annotation process is now about 25% of what it was before the introduction of STM. Finally, the adoption of a robust vocabulary for describing the content of these volumes resulted in a significant increment of the discoverability of relevant publications on SpringerLink, Springer Nature digital library.

Figure 5 shows the average number of yearly downloads for different kinds of SN books published in a specific year. The top line refers to proceedings books in Computer Science before (blue) and after (red) the STM adoption, the intermediate line (grey) refers to the other books in Computer Science, and the yellow line to all the other books. The dotted blue segment branching off the top blue line after 2015 represents the number of downloads for the proceedings books which was to be expected without STM, estimated on the basis of the downloads of other books in Computer Science. The average number of yearly downloads for Computer Science proceedings in Springer Link has doubled since the introduction of STM, increasing from 10K to 20K downloads. This rate of growth compares very favourably with the significantly lower 47% average yearly increase for other book series in Computer Science, which in the same period went up from 7.7K to 11.3K, and with other book series that grew from 6.5K to 10.2K. Hence, considering the expected download in 2016-2018 as a baseline, we estimate that the adoption of STM has resulted in about 9.3 million additional downloads over the last three years. In details, the rate of growth for books annotated with STM increased significantly from 563 (95% Confidence Interval in the range of 106-1019)/year to 2,093 (95%CI 1308-2877)/year ($p<0.05$, with the two 95%CIs not overlapping). The gap in the number of downloads between CS Proceedings and other books in CS had a significant boost from the adoption of STM, jumping from a yearly rate of 324 (95%CI 1-789)/year in the 2004-2015 interval, to 2,092 (95%CI 1308-2878)/year after 2015 ($p<0.05$, with the two 95%CIs not overlapping). In conclusion, while it may not be technically possible to establish a direct causal link between the introduction of STM and the subsequent increase in the number of downloads, all these evidences suggests that STM had a significant positive effect on the number of downloads. This may also indicate that the users are more successful in locating valuable content when it has been annotated with STM.

---

[8] The gold standard is described in [11] and available at https://cso.kmi.open.ac.uk/cso-classifier

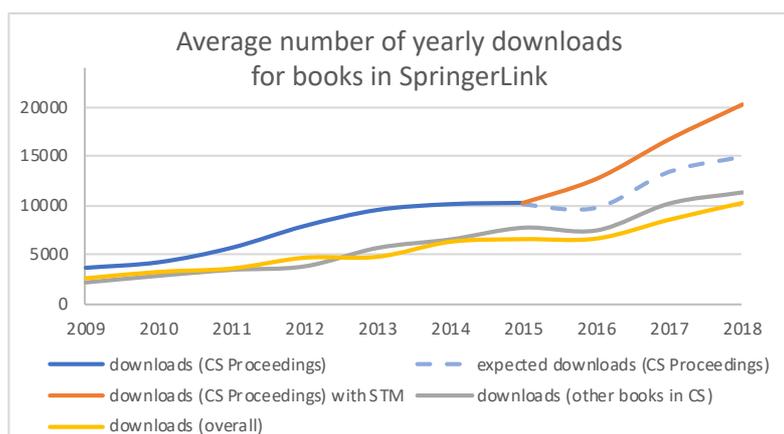

Figure 5. Average number of yearly downloads for books published in a certain year.

## 5 Related Work

In the last years the Semantic Web community produced a vast number of scholarly ontologies (e.g., SWRC[9], BIBO[10], BiDO[11], PROV-O[12], FABIO[13]) and bibliographic repositories in the Linked Data Cloud [17, 18] that can support the analysis of scholarly data. In particular, we saw the emergence of several approaches for linking textual entities to DBpedia, such as DBpedia Spotlight [19], Microsoft Entity Linking[14], BabelFly [20], Illinois Wikifier [21], KORE [22], AGDISTIS [23] and many others. Unfortunately, these systems could not be used for classifying research topics since DBpedia does not offer a particularly good representation of research topics. In particular, it does not contain some of the most recent or specific topics, and it does not structure them in a coherent taxonomy. For instance, the DBpedia entity "Deep Learning" does not currently have any relationship with "Machine Learning" or "Artificial Intelligence".

---

[9] SWRC - http://ontoware.org/swrc/
[10] BIBO - http://bibliontology.com
[11] BiDO - http://purl.org/spar/bido
[12] PROV-O - https://www.w3.org/TR/prov-o
[13] FABIO - http://purl.org/spar/fabio

The task of annotating research papers according to their topics has traditionally been tackled either by using machine-learning classifiers, which assign to the text a number of pre-defined categories, topic models, such as LDA [24], or clustering methods [25, 26]. Here we will focus on the first category, which has the advantage of producing a clean set of formally defined research topics, and thus is usually preferred when it is possible to exploit a vocabulary or taxonomy topics, such as MeSH[15], PhySH[16], or CSO. For instance, Decker [27] introduces an unsupervised approach that generates paper-topic relationships by exploiting keywords and words extracted from the abstracts in order to analyse the trends of topics on different timescales. Mai et al. [28] describe an approach to subject classification which applies deep learning techniques on a training set of scientific papers annotated with the STW Thesaurus for Economics (~5K classes) and MeSH (~27K classes). Shen et al. [29] introduce a technique for concept-document tagging as one of the main components of Microsoft Academic Graph (MAG). Herrera et al. [30] present an approach to categorising the physics literature in terms of the codes from the Physics and Astronomy Classification Scheme (PACS), now replaced by PhySH. Ohniwa et al. [31] perform a similar analysis in the Biomedical domain using the Medical Subject Heading (MeSH). Semantic Scholar[17] is a relevant web application that uses machine language techniques to analyse publications and link them to a number of research areas[18].

Since the Computer Science Ontology is not yet routinely used by researchers, it was not possible to adopt supervised machine learning algorithms that would require a good number of examples for all the relevant categories. For this reason, we opted for an unsupervised solution that does not require such a gold standard.

## 6 Conclusions

In this paper we presented Smart Topic Miner, the tool adopted by Springer Nature editors for annotating all proceedings in Computer Science, and described its evolution over the years, the key lessons learnt, and the impact on the Springer Nature workflow. In particular, we showed how the integration of this ontology-based solution reduced drastically the time used to annotate proceedings and enhanced the discoverability of the relevant publications, resulting in more than 9 million additional downloads.

We plan to further improve STM and to work on different solutions for the automatic production of high-quality metadata describing the content of scientific publications.

---

[14] Microsoft Entity Linking - https://www.microsoft.com/cognitive-services/en-us/entity-linking-intelligence-service
[15] Medical Subject Headings - https://www.nlm.nih.gov/mesh/
[16] Physics Subject Headings - https://physh.aps.org/
[17] Semantic Scholar - www.semanticscholar.org
[18] Topic extraction in Semantic Scholar - https://perma.cc/BP24-WTU7

An important limitation of STM is that it uses exclusively CSO for representing the research topics, and therefore it can only classify publications in Computer Science. For this reason, we are now working on extending STM to consider multiple classification from different fields, starting from Engineering and Natural Sciences. We also intend to investigate methods for learning from the feedback of editors to increase the STM accuracy. Finally, we plan to develop a version of STM to be used by authors for classifying their own research papers when producing their camera ready. We believe that this solution could further improve the quality of the metadata and consequently foster the discoverability of the publications and the diffusion of the relevant scientific ideas.